\documentclass[12pt]{article}
\usepackage{epsfig}

\renewcommand\theequation{\thesection.\arabic{equation}}
\newcommand{\be}{\begin{equation}}
\newcommand{\ee}{\end{equation}}
\newcommand{\bea}{\begin{eqnarray}}
\newcommand{\eea}{\end{eqnarray}}
\newcommand{\bt}{\begin{tabular}}
\newcommand{\et}{\end{tabular}}

\newcommand{\cnb}{{C$\nu$B }}
\newcommand{\pp}{~~~.}
\newcommand{\vv}{~~~,}
\begin{document}

\begin{center}
{\bf NON-THERMAL FEATURES IN THE \\ COSMIC NEUTRINO BACKGROUND}
\end{center}

\begin{center}
{\small \small GIANPIERO MANGANO}
\end{center}

\begin{center}
\noindent {\footnotesize {\it INFN, Sezione di Napoli \\
Dipartimento di Scienze Fisiche, Universit\'{a} di Napoli
"Federico II", \\ Monte
Sant'Angelo, Via Cintia, I-80126 Napoli, Italy} \\
E-mail:mangano@na.infn.it}
\end{center}

\begin{abstract}
{\footnotesize I review some of the basic information on the
Cosmic Neutrino Background momentum distribution. In particular, I
discuss how present data from  several cosmological observables
such as Big Bang Nucleosynthesis, Cosmic Microwave Background and
Large Scale Structure power spectrum constrain possible deviations
from a standard Fermi-Dirac thermal distribution.}
\end{abstract}

\normalsize\baselineskip=15pt

\section{The standard picture}
\setcounter{equation}0

\noindent The large amount of observations accumulated in the last
decades, see \cite{wmap}-\cite{Riess2004} as a highly incomplete
list of references, provided an unprecedented improvement in our
understanding of the general features of the observable Universe.
In fact, this tremendous experimental effort, supplemented by the
generic predictions of inflationary models give a nicely
consistent picture which is nowadays customary to refer to as the
$Standard$ or $Concordance$ Cosmological Model. Despite of the
fact that there are still unsolved major problems, as the nature
of Dark Matter or the so fine tuned value of the cosmological
constant, it is remarkable that the overall picture of the
evolution of the Universe can be described in terms of a
relatively small number of parameters defining the $\Lambda CDM$
model, namely: $^{1)}$ the (cold+hot) dark matter density
$\omega_{\rm dm}=\Omega_{\rm dm} h^2$, $^{2)}$ the baryon density
$\omega_b$, $^{3)}$ the value of the cosmological constant or dark
energy density $\Omega_\Lambda$, $^{4)}$ the normalized value of
the Hubble parameter today $h$, $^{5)}$ the primordial
perturbation spectrum tilt $n_s$ and $^{6)}$ the normalization
$\ln[10^{10} {\cal R}_{\rm rad}]$ where ${\cal R}_{\rm rad}$ is
the curvature perturbation in the radiation era, $^{7)}$ the
optical depth to reionization $\tau$, $^{8)}$ the linear theory
amplitude of matter fluctuations at 8 $h^{-1}$ Mpc $\sigma_8$.

Of course, more exotic scenarios can be considered by adding other
(theoretically motivated) free parameters, such as extra
relativistic degrees of freedom, Quintessence equation of state,
non spatially flat Universe etc, which might help in improving the
agreement of different experimental observations \cite{newwmap}.
In general, the present level of precision of data implies that
several ideas, both theoretically motivated or suggested by
experimental results obtained in different frameworks, can be
tested by checking their implications at the cosmological level.
As an example, the overall scale of neutrino masses $m_0$ can be
bound by comparing data with the expected free streaming
suppression effect of Large Scale Structure (LSS) power spectrum
on scales smaller than $l_{nr}$, the horizon when neutrinos become
non-relativistic \cite{hu,Lesgourgues:2006nd},
\be l_{nr} \sim 38.5 \left(\frac{1~{\rm eV}}{m_0}\right)^{1/2}
\omega_m^{-1/2} ~{\rm Mpc} \vv \label{suppr} \ee
which gives $\sum m_\nu \leq 0.68$ eV \cite{newwmap}, where the
sum is over the three neutrino species. Notice that this bound is
stronger than present constraint from earth-based $^3H$ decay
experiments and of the order of the sensitivity which will be
reached in the near future \cite{katrin}.

It is well known that in the framework of the Hot Big Bang model
we expect the Universe to be filled by a large amount of
neutrinos, with a Fermi-Dirac distribution (in the standard
scenario) characterized by a temperature of 1.95 $^o K$ and
density of 112 cm$^{-3}$/flavor. These relic neutrinos decoupled
from the electromagnetic plasma quite early in time, when weak
interaction rates became slower than the Hubble rate for (photon)
temperatures in the range $2\div 4$ MeV, just before Big Bang
Nucleosynthesis (BBN) took place. Unfortunately, the fact that
this Cosmic Neutrino Background (C$\nu$B) has today a very small
kinetic energy, of the order at most of $10^{-6}$ eV, and that
neutrinos only interact via weak interactions, prevents us from
any possible direct detection of this background on the Earth (see
{\it e.g.} \cite{relicnu,relicnu2} for a recent review).

Nevertheless, there are several indirect ways to constrain the
C$\nu$B by looking at cosmological observables which are
influenced either by the fact that neutrinos contribute to the
Universe expansion rate at all stages, or also via their
interactions with the electromagnetic plasma and baryons before
their decoupling. In this respect, one of the the most sensitive
probe is represented by the values of the light nuclide abundances
produced during BBN. Actually, the final yields of Deuterium,
$^7$Li and in particular of $^4$He strongly depend on the number
of neutrino species as well as on their distribution in phase
space at about 1 MeV when the neutron to proton density ratio
freezes. In fact, since neutrinos were in chemical equilibrium
with the electromagnetic plasma till this epoch we know by
equilibrium thermodynamics that they were distributed according to
a Fermi-Dirac function, yet BBN can constrain exotic features like
the value of their chemical potential
\cite{dolgovetal,Cuoco:2003cu}.

Once decoupled, neutrinos affect all key cosmological observables
which are governed by later stages of the evolution of the
Universe only via their coupling with gravity. A well known
example is their contribution to the total relativistic energy
density, which affects the value of the matter-radiation equality
point, which in turn influences the Cosmic Microwave Background
(CMB) anisotropy spectrum, in particular around the scale of the
first acoustic peak. Similarly, as already mentioned their number
density and their masses are key parameters in the small scale
suppression of the power spectrum of LSS.

The main question which will be addressed in the following
Sections is in my opinion particularly intriguing: how present
(and future) cosmological observations can prove that indeed
neutrinos are thermally distributed? I will first consider the non
thermal features in the \cnb distribution arising from the
neutrino decoupling stage and then discuss the issue on more
general grounds.

\section{Neutrino decoupling}
\setcounter{equation}0 \noindent

Shortly after neutrino decoupling the temperature of the
electromagnetic plasma drops below the electron mass, favoring
$e^\pm$ annihilations that heat the photons. Assuming that this
entropy transfer did not affect the neutrinos because they were
already completely decoupled, it is easy to calculate the
well-known difference between the temperatures of relic photons
and neutrinos $T/T_\nu=(11/4)^{1/3}\simeq 1.40$. However, the
processes of neutrino decoupling and $e^\pm$ annihilations are
sufficiently close in time so that some relic interactions between
$e^\pm$ and neutrinos exist. These relic processes are more
efficient for larger neutrino energies, leading to non-thermal
distortions in the neutrino spectra and a slightly smaller
increase of the comoving photon temperature. These distortions
have been computed by several authors by explicitly solving the
related Boltzmann kinetic equations
\cite{Hannestad:1995rs}-\cite{Mangano:2001iu}, and result to be
very small at the level of few percent, so that their direct
observation is presently out of question. However, they should be
included in any calculation of observables which are influenced by
relic neutrinos. For instance, non-thermal distortions lead to an
enhanced energy density of relic neutrinos parameterized in terms
of the so-called effective number of neutrinos
\cite{Shvartsman:1969mm} $N_{\rm eff}$
\begin{equation}
\rho_{\rm R} = \left[ 1 + \frac{7}{8} \left( \frac{4}{11}
\right)^{4/3} \, N_{\rm eff} \right] \, \rho_\gamma \vv
\label{neff1}
\end{equation}
where $\rho_\gamma$ is the energy density of photons and $\rho_R$
the total radiation energy density. This parameter influences the
CMB anisotropies by shifting the matter-radiation equivalence,
which translates into a different power around the first acoustic
peak via the Integrated Sachs-Wolfe effect. Future CMB
experiments, such as \verb"PLANCK" \cite{planck} or \verb"CMBPOL"
\cite{cmbpol} are foreseen to be sensitive to even a tiny change
in the Universe radiation content, at the level of percent.

As a second main effect, the distortion on neutrino distribution
affects the predictions of BBN. On one hand, the increased value
of $N_{\rm eff}$ shifts the freezing of neutron/proton weak
processes, and so the eventual yield of $^4$He. Furthermore, the
corresponding thermal averaged cross sections for these processes
are also directly affected by any change in the electron neutrino
distribution function. Both effects result into a change of the
$^4$He mass fraction $Y_p$ at the level of $10^{-4}$
\cite{Hannestad:1995rs}-\cite{Esposito:2000hi}. which is quite
small, but it has to be taken into account in precise BBN
numerical codes \cite{Cuoco:2003cu,Serpico:2004gx,cyburt2004}.

Finally, distortions in the neutrino distribution also modifies
the present neutrino number density, so that the contribution of
massive neutrinos to the present energy density of the universe is
also changed.

Recently, the neutrino decoupling stage has been studied taking
into account the effect of flavor oscillations \cite{tegua}. In
this case the neutrino (antineutrino) ensemble is described by $3
{\times} 3$ density matrices
\begin{equation}
\rho(p,t) = \left (\matrix{\rho_{ee} & \rho_{e\mu}&
\rho_{e\tau}\cr \rho_{\mu e}& \rho_{\mu\mu}& \rho_{\mu\tau}\cr
\rho_{\tau e} &\rho_{\tau \mu}& \rho_{\tau \tau} }\right) \pp
\label{3by3}
\end{equation}
The diagonal elements correspond to the usual occupation numbers
of the different flavors, while the off-diagonal terms account for
neutrino mixing.

The equations of motion for the density matrices can be cast in
the form \cite{Sigl:1993fn}
\begin{equation}\label{eq:3by3evol}
i\left(\partial_t-Hp\,\partial_p\right)\rho= \left[ \left(
\frac{M^2}{2p} -\frac{8 \sqrt2 G_{\rm F}\,p}{3 m_{\rm W}^2}{E}
\right),\rho\right] +{C}[\rho] \vv
\end{equation}
where $H$ is the Hubble parameter. The first term in the
commutator corresponds to vacuum oscillation and is proportional
to $M^2$, the mass-squared matrix in the flavor basis, related to
the diagonal one in the mass basis ${\rm diag}(m_1^2,m_2^2,m_3^2)$
via the neutrino mixing matrix. As a reference, in \cite{tegua}
the best-fit values from ref.\ \cite{Maltoni:2004ei} were
considered
\begin{equation}
\left(\frac{\Delta m^2_{21}}{10^{-5}~{\rm eV}^2}, \frac{\Delta
m^2_{31}}{10^{-3}~{\rm eV}^2},\sin^2 \theta_{12}, \sin^2
\theta_{23},\sin^2 \theta_{13}\right)= (8.1, 2.2, 0.3,0.5,0) \vv
\label{oscpardef}
\end{equation}
along with the $3\sigma$ upper bound on $\theta_{13}$, $\sin^2
\theta_{13}=0.047$.

The decoupling of neutrinos takes place at temperatures of the
order MeV, when neutrinos experience both collisions, described by
the collisional integral $C\left[ \rho \right]$ and refractive
effects from the medium. The latter correspond in Eq.\
(\ref{eq:3by3evol}) to the term proportional to the diagonal
matrix $E$, the energy densities of charged leptons.

\begin{center}
\begin{figure}[t]
\includegraphics[width=.7\textwidth]{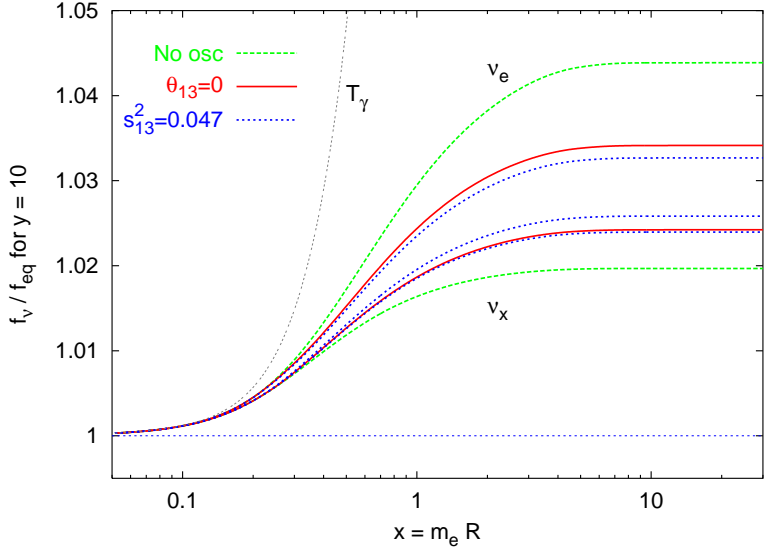}
\caption{\footnotesize \label{fig:evolfnu} Evolution of the
distortion of the $\nu_e$ and $\nu_x=\nu_{\mu,\tau}$ spectrum for
$y=10$. In the case with $\theta_{13}\neq 0$ one can distinguish
the distortions for $\nu_\mu$ (upper line) and $\nu_\tau$ (lower
line). The line labelled with $T_\gamma$ corresponds to the
distribution of a neutrino in full thermal contact with the
electromagnetic plasma. From \cite{tegua}.}
\end{figure}
\end{center}

\begin{center}
\begin{figure}[t]
\includegraphics[width=.7\textwidth]{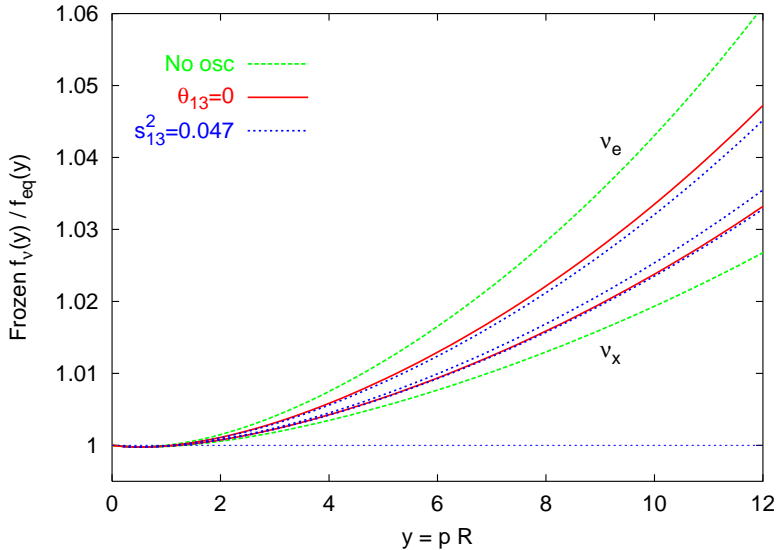}
\caption{\footnotesize \label{fig:finalfnu} Frozen distortions of
the flavor neutrino spectra as a function of the comoving
momentum. In the case with $\theta_{13}\neq 0$ one can distinguish
the distortions for $\nu_\mu$ (upper line) and $\nu_\tau$ (lower
line). From \cite{tegua}.}
\end{figure}
\end{center}

In Fig. \ref{fig:evolfnu} it is shown the evolution of the
distortion of the neutrino distribution as a function of $x=m_e R$
for a particular neutrino comoving momentum ($y =k R=10$), $R$
being the scale factor. At large temperatures or small $x$,
neutrinos are in good thermal contact with $e^{\pm}$. As $x$ grows
weak interactions become less effective in a momentum-dependent
way, leading to distortions in the neutrino spectra which are
larger for $\nu_e$'s than for the other flavors.  Finally, at
larger values of $x$ neutrino decoupling is complete and the
distortions reach their asymptotic values. Fig.\
\ref{fig:finalfnu} shows the asymptotic values of the flavor
neutrino distribution, for the cases without oscillations and with
non-zero mixing. The dependence of the non-thermal distortions in
momentum is well understood and reflects the fact that more
energetic neutrinos were interacting with $e^{\pm}$ for a longer
period. Neutrino oscillations reduce the difference between the
flavor neutrino distortions, and slightly change the final value
of $N_{\rm eff}$ and of the comoving photon temperature $z=T R$.
Both effects translate into a small change of the $^4$He mass
fraction, as also reported in Table 1.

For completeness, I also report the value of the neutrino
distribution function as obtained  by fitting the numerical
results for $\theta_{13}=0$ \cite{tegua}
\begin{eqnarray}
f_{\nu_e}(y) & = & f_{\rm eq}(y) \left[ 1 + 10^{-4}\left(1 - 2.2\,
y + 4.1\, y^2 - 0.047\, y^3\right)\right ]
\nonumber\\
f_{\nu_{\mu,\tau}}(y) & = & f_{\rm eq}(y) \left[ 1 +
10^{-4}\left(-4 + 2.1\, y + 2.4\, y^2 - 0.019\, y^3\right)\right ]
\label{fit_fnualpha}
\end{eqnarray}
These expressions can be easily incorporated into the numerical
tools such as \verb"CMBFAST" \cite{Seljak:1996is} or \verb"CAMB"
\cite{Lewis:1999bs} used to compute CMB and LSS spectra. In fact,
as long as neutrinos are still relativistic the net effect of the
phase space distribution distortion is only via the integrated
effect provided by $N_{\rm eff}$, but a more careful analysis of
effects when neutrinos become non-relativistic should take into
account distortions as a function of neutrino momenta. As an
example, it is easy to calculate from (\ref{fit_fnualpha}) the
present value of neutrino energy density. For the simplest case of
(almost) degenerate neutrinos, {\it i.e.} for the neutrino squared
mass scale $m_0^2$ much larger than the atmospheric squared mass
difference $\Delta m _{31}^2$ one gets
\begin{equation}
\Omega_{\nu} h^2 = \frac{3m_0}{93.14\ ~{\rm eV}} \vv
\label{omeganu}
\end{equation}
where the value in the denominator is slightly smaller than the
analogous result in the instantaneous decoupling limit ($94.12$).
\begin{table*}
\begin{center}
\caption{\small Frozen values of $z$, the neutrino energy density
distortion $\delta\rho_{\nu_\alpha}\equiv
\delta\rho_{\nu_\alpha}/\rho_{\nu_0}$, $N_{\rm eff}$ and $\Delta
Y_p$ including flavor neutrino oscillations. From \cite{tegua}}
%\begin{tabular}{llllllll}
\begin{tabular}{lcccccc}
\hline  & $z$ & $\delta\rho_{\nu_e}$ & $\delta\rho_{\nu_\mu}$
& $\delta\rho_{\nu_\tau}$ & $N_{\rm eff}$ & $\Delta Y_p$\\
\hline $\theta_{13}=0$ & 1.3978 & 0.73\% & 0.52\% & 0.52\% & 3.046
&$2.07 {\times} 10^{-4}$\\ %2.07
$\sin^2\theta_{13}=0.047$ & 1.3978 & 0.70\% & 0.56\% & 0.52\% &
3.046
&$2.12 {\times} 10^{-4}$\\ %2.12
\hline Bimaximal ($\theta_{13}=0$) & 1.3978 & 0.69\% & 0.54\% &
0.54\% & 3.045
&$2.13 {\times} 10^{-4}$\\ %2.07
\hline
\end{tabular}
\end{center}
\label{tab:mixed}
\end{table*}

\section{Bounds on non-thermal features in the \cnb}
\setcounter{equation}0 \noindent

Apart from the non thermal features discussed so far, which are
expected in the framework of Standard Model electroweak
interactions, neutrino distribution might be quite different if
neutrinos also interact with exotic form of matter. Therefore,
obtaining information on the \cnb distribution is a way to
constrain new physics beyond our present knowledge of fundamental
interactions. Thus, it is an interesting issue to understand to
what extent present and future cosmological observations can bound
such exotic interactions by observing the \cnb momentum
distribution.

On completely general ground, we can specify the neutrino
distribution $f_\alpha(y)$ by the set of moments $Q_\alpha^{(n)}$
\cite{clmp}
\be Q_\alpha^{(n)} = \frac{1}{\pi^2}
\left(\frac{4}{11}\right)^{(3+n)/3}T^{3+n} \int y^{2+n}
f_\alpha(y) \, dy \vv \label{momenta} \ee
where the $Q_\alpha^{(n)}$ have been normalized to the standard
value of the neutrino temperature in the instantaneous decoupling
limit $T_\nu= (4/11)^{1/3}T$. Notice that all moments can be
defined if one assumes that neutrino distribution decays at large
comoving momentum as $\exp(-y)$. This is quite expected since at
very high $y$ the shape of the distribution is ruled by the
behavior imprinted by neutrino decoupling as hot relics at the MeV
scale.

If we denote by $P_m(y)$, $m$ being the degree of $P_m(y)$, \be
P_m(y) =  \sum_{k=0}^m c^{(m)}_k y^k \vv \ee the set of
polynomials orthonormal with respect to the measure
$y^2/(\exp(y)+1)$ \be \int_0^\infty dy \,\frac{y^2}{e^y + 1}
P_n(y) P_m(y) = \delta_{nm} \vv \ee then it follows \be
df_\alpha(y) = \frac{y^2}{e^y+1} \sum_{m=0}^\infty F_{\alpha,m}
P_m(y) \, dy \vv \label{series} \ee where \be F_{\alpha,m} =
\sum_{k=0}^m c^{(m)}_k Q_\alpha^{(k)} T_\nu^{-k} \vv \ee i.e. a
linear combination of moments up to order $m$ with coefficients
$c^{(m)}_k$.

For a Fermi-Dirac distribution all moments can be expressed in
terms of the number density $Q_\alpha^{(0)}$ or, equivalently, as
functions of the only independent parameter $T_\nu$, the first two
moments being related to $N_{\rm eff}$ and $\Omega_\nu$ today,
respectively
\be \omega_{\nu} = \Omega_{\nu} h^2 = 0.058 \, \frac{m_0}{{\rm
eV}} \frac{11}{4}T^{-3} Q^{(0)}_\alpha \vv \label{omnu} \ee
\be N_{\rm eff} = \frac{120}{7\pi^2}
\left(\frac{11}{4}\right)^{4/3} T^{-4}\sum_\alpha Q_\alpha^{(1)}
\vv \label{neff} \ee
where we have assumed for simplicity that the three neutrinos
share the same distribution.

In the following I consider only the first two moments
$\omega_{\nu}$ and $N_{\rm eff}$ as free and independent
parameters, to be constrained using cosmological data \cite{clmp}.
In fact, the only way to decide how many $Q^{(n)}_\alpha$ should
be included in the analysis can be only dictated by the
sensitivity of the available observational data to the distortion
of the neutrino distribution. Presently (and surprisingly), it is
already very hard to get quite strong constraints on the first two
moments. In case future observations would reach a higher
sensitivity on neutrino distribution, it would be desirable to
include higher order moments, such as the skewness or the
kurtosis, related to $Q_\alpha^{(2)}$ and $Q_\alpha^{(3)}$,
respectively.

For the sake of definiteness in \cite{clmp} a specific example was
worked out in details, where neutrinos interact with a light
scalar field $\Phi$ with mass $M$ via the interaction lagrangian
density
\be {\cal L}_{int}= \frac{\lambda}{\sqrt{3}} \Phi \sum_i
\overline{\nu}_i\nu_i \label{lagr} \pp \ee
This majoron-inspired model have been also considered in
\cite{hannestad1}-\cite{dodelson} for a very small light $\Phi$
field, even lighter than neutrinos. In this case, neutrino
annihilations would results into a $neutrinoless$ cosmological
model \cite{dodelson}. On the other hand, for $m_\Phi\geq 2 m_0$
the decays of the unstable $\Phi$ particles might lead to non
trivial features in neutrino distribution provided
\begin{itemize}
\item[i)] decays take place out of equilibrium, {\it i.e.} for
temperatures smaller than the decaying particle mass $M$.
\item[ii)] occur after weak interaction freeze-out, otherwise
neutrino interactions with the electromagnetic plasma would erase
any non thermal features.
\end{itemize}
Of course, different scenarios can be considered, as for example
the case of unstable neutrinos $\nu_h$ decaying into a (pseudo)
scalar particle $\Phi$ and a lighter neutrino $\nu_l$,
\be \nu_h \rightarrow \Phi \, \nu_l \vv \ee
but I notice that at this stage the main issue is rather to
understand how present (and future) data can constrain the
non-thermal contribution to the neutrino background, rather than
to discriminate among different models. Despite of the fact that
in \cite{clmp} the out of equilibrium $\Phi$ decay case was taken
as the reference model, all results are quite general and can be
applied to different scenarios as well.
\begin{center}
\begin{figure}
\includegraphics[width=.7\textwidth]{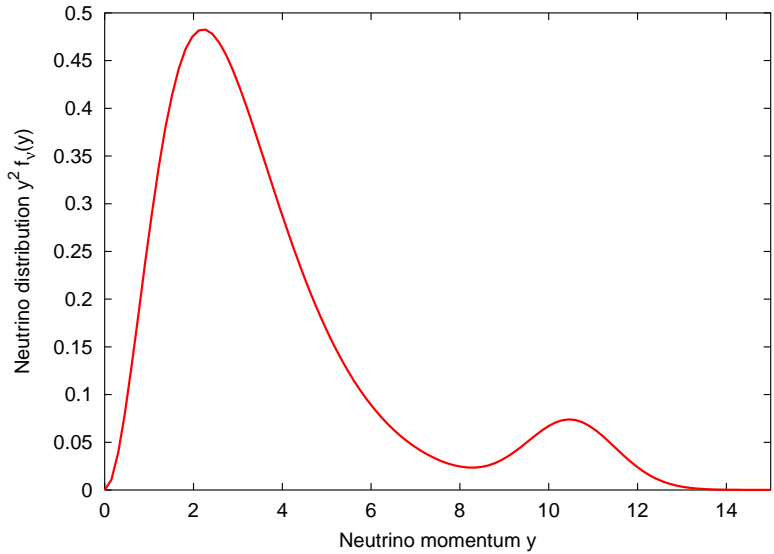}
\caption{\footnotesize \label{fig:NT} Differential number density
of relic neutrinos as a function of the comoving momentum for the
non-thermal spectrum in Equation (\ref{NTspec}). The parameters
are $A=0.018$, $y_*=10.5$ and $\sigma=1$, which corresponds to
$N_{\rm eff} \simeq 4$. From \cite{clmp}.}
\end{figure}
\end{center}
When the $\Phi$ particles decay, the neutrino distribution gets an
additional contribution, which in the narrow width limit and in
the instantaneous decay approximation at temperature $T_D$
corresponds to a peaked pulse at $y_*=M/(2 T_D)$ so that
\be y^2 f(y) dy = y^2 \frac{1}{e^y+1} dy \nonumber +\pi^2
\frac{A}{\sqrt{2 \pi \sigma^2}} \exp \left[-\left(\frac{y-y_*}{2
\sigma^2}\right)^2\right] \pp \label{NTspec} \ee
An example of this non-thermal neutrino spectrum is shown in
Figure \ref{fig:NT}. The parameter $A$ is given by the comoving
$\Phi$ number density at decay $A= n_\Phi(T_D) R^3$

\begin{figure}
\includegraphics[width=.7\textwidth]{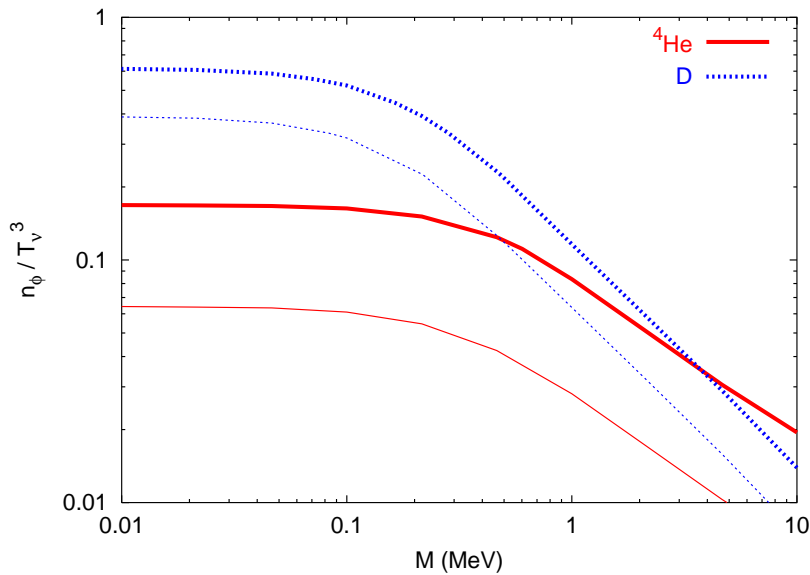}
\caption{\footnotesize \label{fig:BBNhe} The 1 $\sigma$ (thin
lines) and 2 $\sigma$ (thick lines) BBN bounds on the $\Phi$
number density (normalized to $T_\nu^3$) versus mass $M$ in MeV.
The regions above the contours would be in disagreement with the
observed primordial abundances of $^4$He or D. From \cite{clmp}.}
\end{figure}

Correspondingly, the lower moments expressed in terms of the
parameters of Eqs. (\ref{neff}) and (\ref{omnu}) read
\begin{eqnarray}
\omega_{\nu} &=& \frac{m_0}{93.2 ~{\rm eV}} \left(1+ 0.99 \frac{2
\pi^2}{3
\zeta(3)}\,A \right) \vv \label{nonthomnu} \\
N_{\rm eff} &=& 3.04 \left(1+ 0.99 \frac{120}{7\pi^2}\,A y_*
\right) \pp \label{nonthneff}
\end{eqnarray}
For sufficiently weakly interacting $\Phi$ particles, so that they
are decoupled from the thermal bath since at least the BBN epoch,
the largest value of $A$ can be bound by BBN as a function of the
$\Phi$ mass and number density. Large values of $A$ and $M$ in
fact, implies a large value of the $\Phi$ energy density during
BBN, thus affecting the Hubble expansion rate and the final
abundances of $^4$He and Deuterium. Using present data on these
two nuclei, see {\it e.g. } \cite{Serpico:2004gx}, this translates
into $A \leq 0.1$, unless the value of $M$ is order MeV or larger,
in which case $A$ is much more constrained, see Fig.
\ref{fig:BBNhe}.

The model described so far can also be tested using CMB and LSS
power spectrum. The results are summarized in third column of
Table \ref{ranges}. For comparison I show also the corresponding
result for the case where neutrino distribution is assumed to be
the standard Fermi-Dirac, but the value of $N_{\rm eff}$
corresponding to extra relativistic degrees of freedom is taken as
a free parameter (second column). In particular, the best value of
the effective $\chi^2$ for the three models is shown in the first
line. Notice that both models do not improve significantly the
$\chi^2$ with respect to the standard $\Lambda$CDM, despite of
their larger parameter space.
\begin{table}[t]
\begin{tabular}{lccc}
\hline \hline &$\Lambda$CDM&$\Lambda$CDM + extra
radiation&$\Lambda$CDM+non thermal $\nu$
\\
\hline $\chi^2_{\rm min}$&1688.2&1688.0&1688.0
\\
\hline
$\ln[10^{10} {\cal R}_{\rm rad}]$&$3.2{\pm}0.1$&$3.2{\pm}0.1$&$3.2{\pm}0.1$\\
$n_s$&$0.97{\pm}0.02$&$0.99{\pm}0.03$&$1.00{\pm}0.03$\\
$\omega_b$&$0.0235{\pm}0.0010$&$0.0231{\pm}0.0010$&$0.0233{\pm}0.0011$\\
$\omega_{\rm
dm}$&$0.121{\pm}0.005$&$0.17{\pm}0.03$&$0.17{\pm}0.03$\\
$\theta$&$1.043{\pm}0.005$&$1.033{\pm}0.006$&$1.033{\pm}0.006$\\
$\tau$&$0.13{\pm}0.05$&$0.13{\pm}0.06$&$0.15{\pm}0.07$\\
$\beta$&$0.46{\pm}0.04$&$0.48{\pm}0.04$&$0.48{\pm}0.04$\\ $m_0$
(eV)&$0.3{\pm}0.2$&$0.8{\pm}0.5$&$0.7{\pm}0.4$\\ $N_{\rm eff}$&3.04&$6{\pm}2$&$6{\pm}2$\\
$q$&1&1&$1.25{\pm}0.13$\\
\hline
$h$&$0.67{\pm}0.02$&$0.76{\pm}0.06$&$0.76{\pm}0.05$\\
Age (Gyr)&$13.8{\pm}0.2$&$12.1{\pm}0.9$&$12.1{\pm}0.8$\\
$\Omega_{\Lambda}$&$0.68{\pm}0.03$&$0.67{\pm}0.03$&$0.67{\pm}0.03$\\
$z_{re}$&$14{\pm}4$&$16{\pm}5$&$18{\pm}6$\\
$\sigma_8$&$0.76{\pm}0.06$&$0.77{\pm}0.07$&$0.77{\pm}0.07$\\
\hline \hline
\end{tabular}
\caption{ Minimum value of the effective $\chi^2$ (defined as
$-2\ln{\cal L}$, where ${\cal L}$ is the likelihood function) and
1$\sigma$ confidence limits for the parameters of the three models
under consideration. $\theta$ is the ratio of the sound horizon to
the angular diameter distance multiplied by 100, while $\beta$ is
the 2dF redshift-space distortion factor. Finally, $q \equiv
\omega_{\nu} (93.2\,\mathrm{eV} / m_0)$. The last five lines refer
to parameters which can be obtained by the set of independent
parameters of the first ten lines. From \cite{clmp}.
\label{ranges}}
\end{table}
Notice that the bound on the parameter $q \equiv \omega_{\nu}
(93.2\,\mathrm{eV} / m_0)$ comes essentially from the BBN prior
$A<0.1$, since the CMB and LSS data alone would be compatible with
much larger deviations from a thermal phase-space distribution (up
to $A=1$ at 2-$\sigma$). Similarly, $y_*$ is poorly constrained,
and large values for it are still allowed. In the $\Phi$ decay
scenario this implies that these decays can take place in highly
out of equilibrium conditions, $T_D \ll M$, the only bound being
instead on the scalar particle number density at the BBN epoch.

The fact that $A$ and $y_*$ are so poorly constrained is due the
existence of a degeneracy direction involving mainly $\omega_{\rm
dm}$, $m_0$ and $N_{\rm eff}$ for the two models in which the
radiation density near the time of decoupling is a free parameter.
Values as large as $\omega_{\rm dm}=0.23$, $m_0 = 1.5$~eV or
$N_{\rm eff}=9$ are still allowed at the 2-$\sigma$ level.
\begin{table*}
\begin{tabular}{lccccccccr}
\hline \hline name  &$n_s$ &$\omega_b$ &$\omega_m$
&$\Omega_{\Lambda}$ &$\tau$  &$m_0$ (eV) &$N_{\rm eff}$ &$q$ &
\\
\hline fiducial values  &0.96 &0.023 &0.14 &0.70 &0.11 &0.5 &4.0
&1.1 &
\\
\hline 1$\sigma$ error  &0.009 &0.0003 &0.004 &0.02 &0.005 &0.3
&0.1 &0.7 &
\\
\hline \hline
\end{tabular}
\caption{Expected errors on the parameters of the $\Lambda$CDM +
non thermal neutrino model. The first line shows the fiducial
values, i.e. the parameter values assumed to represent the best
fit to the data. The second line give the forecast for the
associated 1$\sigma$ errors for PLANCK + SDSS.}
\end{table*}
An improvement is expected in removing these degeneracies from
future experiments, though not particularly dramatic. In Table 3
it is shown the result of a standard Fisher matrix analysis
assuming a fiducial model as reported in the first line of the
Table. The forecast is based on the foreseen sensitivity of
\verb"PLANCK" combined with the completed SDSS redshift survey
with effective volume $V_{\rm eff}=1\,h^{-3}$ Gpc$^3$ and a free
bias. While the sensitivity on $N_{\rm eff}$ is now at the level
(or better) than percent, both $q$ and $m_0$ are not very well
constrained, since these two parameters are measurable only from
the free-streaming effect in the matter power spectrum, while all
other parameters have a clear signature in the CMB anisotropies.

\section{Conclusions}
The role of \cnb in cosmology is quite ubiquitous, yet it is quite
disappointing that any $direct$ detection of background neutrinos
will be very hard to achieve. Through their influence on several
cosmological observables several properties of neutrinos can be
already constrained, such as their mass, possible finite
lifetimes, magnetic moments, etc. On the other hand, possible
exotic features in their momentum distribution would also
represent quite a unique imprint of new interactions beyond our
present understanding of fundamental physics. Unfortunately,
present data have not already reached enough sensitivity to put
severe constraints in this respect, despite of the huge
improvements in the last years, but this sensitivity is likely to
be highly improved in the near future. Waiting for these new
exciting times it is worth scrutinizing new theoretical
perspectives, keeping in mind as a general warning what Pauli said
to his friend W. Baade about his neutrino hypothesis \cite{hoyle}
(see D. Haidt
in these Proceedings)\\

{\it "... Today I have done something which no theoretical
physicist should ever do: I have predicted something which shall
never be detected experimentally..."}

\section{Acknowledgments}
I warmly thank A. Cuoco,  J. Lesgourgues, G. Miele, S. Pastor, M.
Peloso, O. Pisanti, T. Pinto and P.D. Serpico for all discussions
and fruitful collaboration on the topics summarized in this paper.

\end{document}